\documentclass[12pt]{article}
\usepackage{graphicx}
\begin{document}

\pagestyle{empty}

\noindent
{\bf Studies of the Long Secondary Periods in Pulsating Red Giants}

\bigskip

\noindent
{\bf John R. Percy and Emily Deibert\\Department of Astronomy and Astrophysics, and\\Dunlap Institute of Astronomy and Astrophysics\\University of Toronto\\Toronto ON\\Canada M5S 3H4}

\bigskip

{\bf Abstract}  We have used systematic, sustained visual observations from
the AAVSO International Database, and the AAVSO time-series analysis
package VSTAR to study the unexplained ``long secondary periods" (LSPs) in 27 pulsating
red giants.  In our sample, the LSPs range from 479 to 2967 days, and are on average 8.1 $\pm$ 1.3  
times the excited pulsation period; they are about 5 times the fundamental period.  There is no evidence for more than one LSP in
each star.  In stars with both the fundamental and first overtone radial
period present, LSP is more often about 10 times the latter.  The visual amplitudes of the LSPs are typically 0.1 magnitude
and do not correlate with the LSP.  The phase curves tend to be sinusoidal,
but at least two are sawtooth.  The LSPs are stable, within their errors, over
the timespan of our data, which is typically 25,000 days.  The amplitudes, however, vary by up to a factor of two or more on a time scale
of roughly 20-30 LSPs.  There is no obvious difference between the behavior
of the carbon (C) stars and the normal oxygen (M) stars.  Previous multicolor
photoelectric observations showed that the LSP color variations are similar
to those of the pulsation period, and of the LSPs in the Magellanic
Clouds, and not like those of eclipsing stars.
We note that the LSPs are similar to the estimated rotation periods of the
stars, though the latter have large uncertainties.  This suggests that the
LSP phenomenon may be a form of modulated rotational variability.

\medskip

\noindent
AAVSO keywords = AAVSO International Database; Photometry, visual; pulsating variables; giants, red; period analysis; amplitude analysis

\medskip

\noindent
ADS keywords = stars; stars: late-type; techniques: photometric; methods: statistical; stars: variable; stars: oscillations

\medskip

\noindent
{\bf 1. Introduction}

\smallskip

About a third of pulsating red giants show a long secondary period (LSP),
about ten times longer than the pulsation period (Wood 2000, Percy and Bakos 2003).  LSPs have been known
for many decades (O'Connell 1933, Payne-Gaposchkin 1954, Houk 1963), from
visual or photographic observations.  LSPs
have also been discovered photoelectrically in many small-amplitude pulsating red giants
(Percy {\it et al.} 1996, Percy {\it et al.} 2001, Percy and Bakos 2003).  
LSPs
were rediscovered in 1999 by
Peter Wood and colleagues (Wood {\it et al.} 1999, Wood 2000), from large-scale survey observations of pulsating
red giants in the Large Magellanic Cloud (LMC).
Since then, he and others have accumulated many more observations
of this phenomenon, and have considered many possible explanations.
None of them explains these LSPs satisfactorily.  Wood considers this to be the most
significant unsolved problem in stellar pulsation.

There are at least three possible mechanisms that are still of interest: the turnover of
giant convective cells (Stothers 2010), oscillatory convective modes (Saio
{\it et al.} 2015), or the presence of a dusty cloud, orbiting the red
giant, together with a low-mass companion in a close, circular orbit
(Soszy\'{n}ski and Udalski 2014), or some combination of these.

Nicholls {\it et al.} (2009) (hereinafter NWCS) discussed the problem of
the LSPs in detail, and ``are unable to find a suitable model for the
LSPs", so they ended by listing ``all the currently known properties
of LSPs" -- a list of 13 items.  A few other properties have been added
in the literature since then.  The purpose of the present paper is to
use the systematic, sustained visual data in the AAVSO International
Database (Kafka 2016) to add further ``knowns" to this list.

LSPs are not the only unsolved mystery in pulsating red giants.  Percy and Abachi (2013)
found that the pulsation amplitudes in these stars varied by factors of 2 to
over 10 on median time scales of 44 pulsation periods.
There are also random cycle-to-cycle
period fluctuations in red giants (Eddington and Plakidis 1929, Percy and Colivas 1999) which may
be related to large convection cells in these stars.
 
\medskip

\noindent
{\bf 2. Data and Analysis}

\smallskip

We used visual observations from the AAVSO International Database (AID:
Kafka 2016), and the AAVSO VSTAR time-series analysis package (Benn 2013),
which includes both a Fourier analysis and a wavelet analysis routine.
Stars (listed in Table 1) with sufficient observations in the AID were
chosen for analysis from several sources, including Payne-Gaposhkin (1954), Houk (1963),
and Kiss {\it et al.} (1999), among others. We have included three pulsating
red supergiants (SG) for comparison.  We have also included V Hya which has 
unusually-deep and stable LSP minima, and may be an eclipsing binary star (Knapp
{\it et al.} 1999).  The columns give the star name, the
type (M, C, or SG), the starting JD (otherwise all the AID data were used) the pulsation period P, the LSP, the ratio of these,
the mean, maximum, and minimum LSP amplitudes, and notes.  The notes are as
follows: pcs - sinusoidal phase curve; pcs: - possibly sinusoidal phase
curve; pc? - phase curve uncertain, but possibly non-sinusoidal; pcst - phase
curve sawtooth; dsp - data sparse; * - see section 3.12. The phase curve is a
graph of magnitude versus phase, determined with VSTAR using the known period;
it essentially folds all of the observations into one cycle.  Note that the mean
amplitude was determined by Fourier analysis, the maximum and minimum
amplitude by wavelet analysis.

Figure 1 shows the long-term
light curve of U Del.  The visual observations have also been averaged in
bins of 119 days (the pulsation period) to show the LSP more clearly.  

\medskip

\noindent
{\bf 3 Results}

\smallskip

\medskip

\noindent
{\bf 3.1 Lengths of LSPs}

\smallskip

NWCS gave a range of 250 to 1400 days, primarily based on observations of
stars in the Magellanic Clouds.  Our study includes both short-period and
longer-period stars in the Milky Way.  The Galactic stars in Table 1 may
have different properties than the low-metallicity stars in the Magellanic
Clouds, studied by Wood and others.  Excluding supergiants, our range is
479 to 2967 days.  There may be stars with shorter or longer LSPs, presumably
with pulsation periods shorter or longer than those in Table 1.  

Figure 2 shows the relationship between the LSP and the excited
pulsation period.  The slope of the relationship is 8.1 $\pm$ 1.3. The
relationship may not be linear.  The results for supergiants (section 3.2)
suggest that the slope becomes shallower at longer periods.  This may be
because the fundamental period is more likely to be excited in longer-period
stars; see Section 3.5. 

\medskip

\noindent
{\bf 3.2 LSPs in Red Supergiants}

\smallskip

It is important to note that LSPs are also found in pulsating red
{\it supergiants} (Kiss {\it et al.} 2006, Percy and Sato 2009);
the LSP
phenomenon seems to be continuous from class III to class II to class I stars.
Kiss {\it et al.} (2006) used Fourier techniques to analyze visual measurements;
Percy and Sato (2009) used self-correlation techniques to analyze similar 
datasets.   There are about 10 stars for which the above-mentioned
papers obtained consistent pulsation periods and LSPs; for these, the median
value
of LSP/P is 6.0 and the median LSP visual amplitude is 0.10 mag, similar to those
in red giants.  For red supergiants
in the Large Magellanic Cloud, the median LSP/P is about 4, and the
median V amplitude is about 0.08 (Yang and Jiang 2012).  Kiss {\it et al.} (2006) concluded, on the basis of the Lorentzian envelopes of the peaks in the
Fourier spectra, and the strong 1/f noise, that large convection cells
play an important role in the behavior of these red supergiant stars.

\medskip

\noindent
{\bf 3.3 The Amplitudes of the LSPs}

\smallskip

The median LSP amplitude for the stars in Table 1 is 0.10 for both the M
stars and the C stars.  There is no evidence that it varies with LSP (and
therefore with the size and/or temperature of the star), according to
Figure 3, which shows the LSP amplitude as a function of LSP. Note that we
find LSPs in short-period pulsators such as RZ Ari, whose variability is
best studied photoelectrically (Percy {\it et al.} 2008).  Nicholls {\it et al.} (2010)
show that the LSP {\it velocity} amplitude is constant, at a few km/s, over
a very wide range of LSPs. 

It is said
that Mira stars do not have LSPs, but low-amplitude LSPs may be hidden
by the large-amplitude pulsation, and by the complex Fourier spectra of these
stars, which arise from the systematics in the time distribution of the
observations, and from the stars' random changes in period and amplitude.
For instance: the Fourier spectra of AAVSO visual data on Mira stars
tend to have strong one-cycle-per-year aliases at periods of a few thousand
days (the possible values of the LSPs), and the noise level in the spectra
exceeds 0.1 magnitude.  

\medskip

\noindent
{\bf 3.4 The Phase Curves of the LSPs}

\smallskip

For most of the stars in our sample, the LSP phase curves are indistinguishable
from sine curves, with the exception of a very small number which have
sawtooth or possibly sawtooth shapes.  The clearest example is Y Lyn 
(Figure 4).  This figure shows the long-term light curve of Y Lyn; the
visual observations have also been averaged in bins of 134.7 days (the
pulsation period) to show the sawtooth shape of the LSP light curve more clearly.  The other star with a distinctly saw-tooth phase curve is RV Lac.  We note
that both these stars have higher-than-average LSP amplitudes.  The model of Soszy\'{n}ski and Udalski (2014) predicts a different shape for
the phase curve; see section 4.

\medskip

\noindent
{\bf 3.5 Is the LSP Always Ten Times the Fundamental Pulsation Period?}

\smallskip

Some stars with LSPs pulsate in the fundamental mode, and others in the
first overtone.  Is the LSP always ten times the excited period, no matter
which mode is excited?

Fuentes-Morales and Vogt (2014) used the ASAS database to study 72 pulsating
red giants.  Of the stars which they identified as triply periodic (their
Table 2), eight appear to have an LSP and a fundamental (P0) and a first
overtone (P1) radial period present.  In three of these, LSP/P1 is closer to 8.1 
than LSP/P0; in two of these, LSP/P0 is closer, and in three of them, both
ratios are equidistant from 8.1.  In Kiss {\it et al.} (1999)'s list of
triply-periodic stars, there are nine with LSP, P0 and P1.  In six cases,
LSP/P1 is closer to 8.1, in two, LSP/P0 is closer,  and in one case, they
are equidistant from 8.1.
In our sample, there are
four stars (RX Boo, RS Cam, TX Dra, and X Her) which appear to
have these three periods present.  In three, LSP/P0 is closer to 8.1; in the
other, LSP/P1 is closer to 8.1.  We conclude that, although LSP is slightly more
often about 8.1 times the first overtone radial period, this is not always
the case.  This may explain some of the scatter in Figure 2.

A different approach is to consider a plot (Figure 5) of LSP/P0 and LSP/P1 versus LSP.
For stars with LSP less than about 1500 days, LSP/P0 is about 5-6, LSP/P1 is
about 9-11.  This suggests that the LSP is generally about 5-6 times the
fundamental period.  For stars with LSP greater than 1500 days, the behavior is much
less consistent.

\medskip

\noindent
{\bf 3.6 Uniqueness of the LSP}

\smallskip

Some theories of the LSP suggest that there may be two or more long timescales
in a pulsating red giant, such as the turnover time of a bright or dark convective cell, and
the rotation period of the star.
There are already four timescales which we have identified in our sample:
(1) the pulsation period; (2) the LSP; (3) the timescale for increase and
decrease of the amplitude of the pulsation period; and (4) the timescale
for increase and decrease of the amplitude of the LSP.  As noted below,
timescale (3) is not the same as the LSP.

We have used Fourier analysis to look for stars in our sample which might have more than one LSP,
but there are none in which a second LSP is significantly above the noise
level.  For example: the LSP light curves in Figures 1 and 4 appear to be monoperiodic.

\medskip

\noindent
{\bf 3.7 LSP Period Stability}

\smallskip

For most of the stars in Table 1 -- those for which the observations are
both dense and sustained -- the half-width at half-maximum errors in determining the LSP are 3 to 5 percent.  Figure 6 shows the Fourier spectrum of TZ And,
showing how the HWHM uncertainly of the LSP was determined.  We adopt 3 percent as a reasonable intrinsic
uncertainty caused by the finite length of the dataset and the scatter in the
visual observations.
We then used wavelet analysis to study changes in the LSP with time.  We
did not find any stars for which LSP varied by more than three times the
error i.e. by 10 percent or more, over the timespan of the observations,
which is typically 25,000 days.  
A possible exception may be Z Eri, for which
the mean LSP is 722 days, the range is 695 to 740 days, and the formal error of the LSP is
only 7 days.  This error seems unreasonably low, however, being only 1 percent.

\medskip

\medskip

\noindent
{\bf 3.8 LSP Amplitude Stability}

\smallskip

Percy and Abachi (2013) showed that, for a small sample of red giants with LSPs,
the amplitudes of most of the LSPs varied by a factor of up to 2, on a time scale
of typically 30 LSPs or greater.  Figure 7 shows the amplitude of the LSP of
RT Cnc as a function of time.  The amplitude rises and falls on a time scale
of about 20 LSPs.  This is true of the other stars in Table 1, as it was for
the LSPs of the stars studued by Percy and Abachi (2013).  It is difficult
to determine the time scale of the LSP amplitude variability because of its
length, but one way of doing so is shown in Figure 8, which plots a measure
of the {\it change} in LSP amplitude against the LSP.  Stars with longer LSPs change
less in amplitude, presumably because, for these stars, the time scale of amplitude variability, being
tens of times greater than the LSP itself, is longer than the
timespan of our observations.  This interpretation of Figure 8 is therefore consistent
with the results of Percy and Abachi (2013).

\medskip

\noindent
{\bf 3.9 Correlation of Pulsation Amplitude with LSP Phase}

\smallskip

NWCS note that the primary pulsation is visible in the light curve at all
times throughout the LSP and the primary {\it period} does not significantly
change with LSP phase.  Because the cycles of increase and decrease of
the primary pulsation period's {\it amplitude} vary on a median time scale of 44 
periods (Percy and Abachi 2013), whereas the LSP is about 8.1 times the excited pulsation period, it follows that
the pulsation period's {\it amplitude} is not correlated with the LSP phase.  For
instance: for W Ori, the cycles of pulsation amplitude increase and decrease are
about 4500 days, whereas the LSP is about half that.

\medskip

\noindent
{\bf 3.10 Oxygen (M) Stars versus Carbon (C) Stars}

\smallskip

In our previous studies of red giants, Percy and Yook (2014) identified one
possible difference in pulsational behavior between oxygen (M) stars and
carbon (C) stars, namely that, in C stars, amplitude changes correlated with
period changes, whereas this was not true for M stars.
We have therefore looked in our database to see whether there are any
differences between M and C stars in the LSPs or their amplitudes.  In doing this, we have been
careful to compare stars with similar periods.  Carbon stars have longer
pulsation periods, on average, than oxygen stars, because the carbon-star
phenomenon is associated with a more advanced stage of evolution, when the
star is larger and cooler.  Figures 2, 8, and 9 show no obvious differences
between the behavior of the two classes of star.

\medskip

\noindent
{\bf 3.11 LSP Color to Light Variations}

\smallskip

Derekas {\it et al.} (2006) pointed out that, on the basis of their observations, the relative
color amplitudes were similar, for the LSPs, to those for the pulsation period,
but this was not true for stars with ellipsoidal variability.
Percy {\it et al.} (2008) used merged AAVSO and robotic telescope photometry
to show that the same was true for the LSPs of 12 small-amplitude pulsating red giants (EG And, RZ Ari, Psi Aur, BC CMi, TU CVn, FS Com, $\alpha$ Her, V642 Her,  30 Her, Y Lyn, UX Lyn, TV Psc),
with the possible exception of EG And, which appears to be a genuine symbiotic
spectroscopic binary (their figures 7 and 8).  Specifically: these stars had
a visual amplitude about twice the red amplitude which, in turn, was about
twice the near-infrared amplitude; they are brighter when hotter.  These relative
amplitudes are slightly larger, on average, than the $\Delta$V/$\Delta$I
ratios shown in Figure 4 of Soszy\'{n}ski and Udalski (2014), but are within
the same range.  And the metallicities of the Galactic and LMC stars are
quite different.

\medskip

\noindent
{\bf 3.12 Notes on Individual Stars}

\smallskip

{\it T Ari:} This star, because of its large amplitude, is classified as a Mira star.

{\it RX Boo:} There may also be a 372.99-day period.

{\it Y CVn:} The pulsation period is uncertain, but is most likely 160 days.

{\it AW Cyg:} The pulsation period (most likely 358 days) and the LSP are
uncertain.

{\it RY Dra:} The 276.7-day pulsation period is consistent with the DIRBE
photometry (Price {\it et al.} 2010), and the 1135.6-day LSP in the AAVSO data seems secure.  But the
DIRBE data suggests that the LSP may be longer.

{\it g Her:} There may also be a 61.21-day pulsation period.

{\it V Hya:} This is a peculiar pulsating red giant with an LSP (6907 days) which has a large amplitude (1.13 mag), a V-shaped phase curve, and a stable LSP
and LSP amplitude -- very much like the RVB star U Mon.  Knapp {\it et al.} 1999) suggest that this is a binary in which the secondary component is surrounded by dust.

{\it $\tau$4 Ser:} The pulsation period is uncertain, but is most likely
111 days.

{\it V UMi:} There may also be a 125.45-day pulsation period.

\medskip
 
\smallskip

\begin{figure}
\begin{center}
\includegraphics[height=7cm]{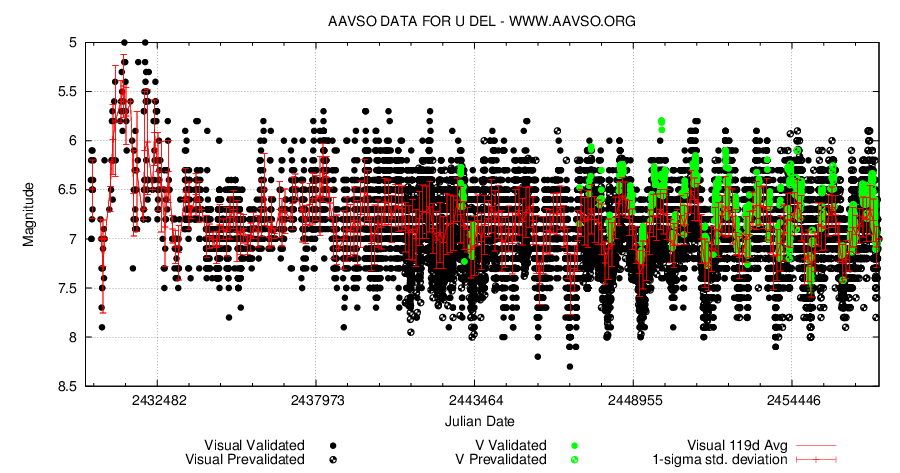}
\end{center}
\caption{The long-term light curve of U Del, also showing the data averaged
in bins of 119 days (the pulsation period) to shown the LSP light curve (red curve) more clearly.  The LSP, as determined by VSTAR, is 1162.8 days, and this is consistent
with the value which would be determined by counting cycles in this figure. We
have plotted all the observations, to show how the density of observations
and the clarity of the LSP increase after JD 2440000.  The black points are
visual; the green ones are photoelectric V.}
\end{figure}

\begin{figure}
\begin{center}
\includegraphics[height=7cm]{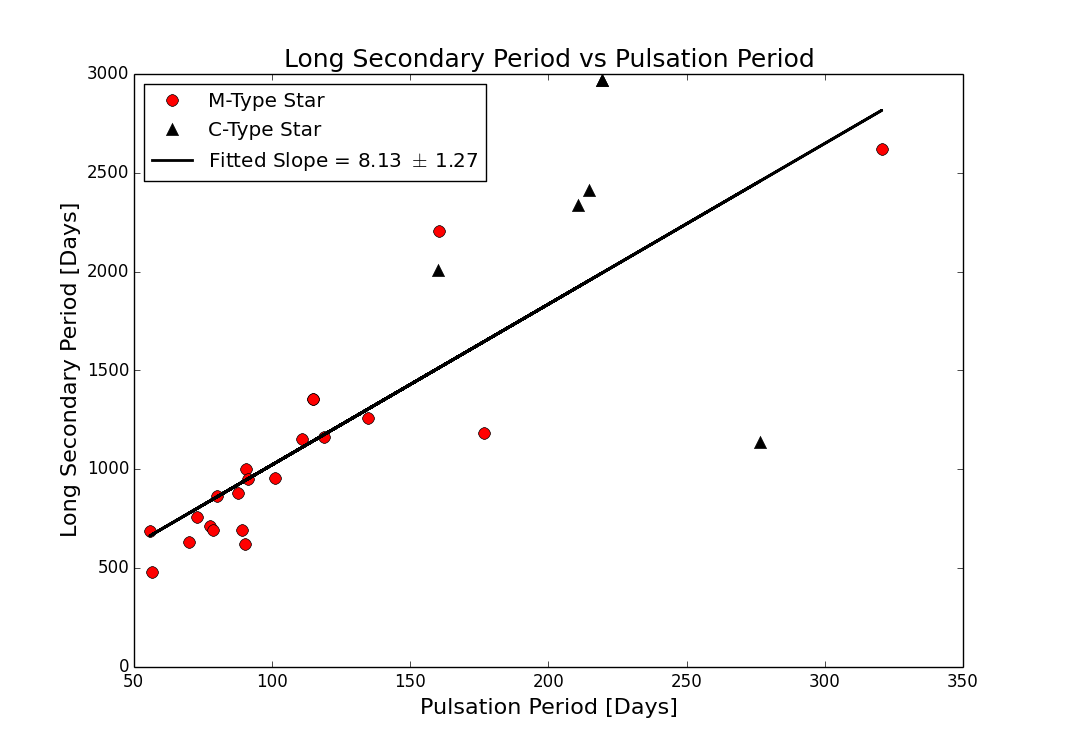}
\end{center}
\caption{The LSP as a function of pulsation period.  The slope of a linear fit is 8.1 $\pm$ 1.3.}
\end{figure}

\begin{figure}
\begin{center}
\includegraphics[height=7cm]{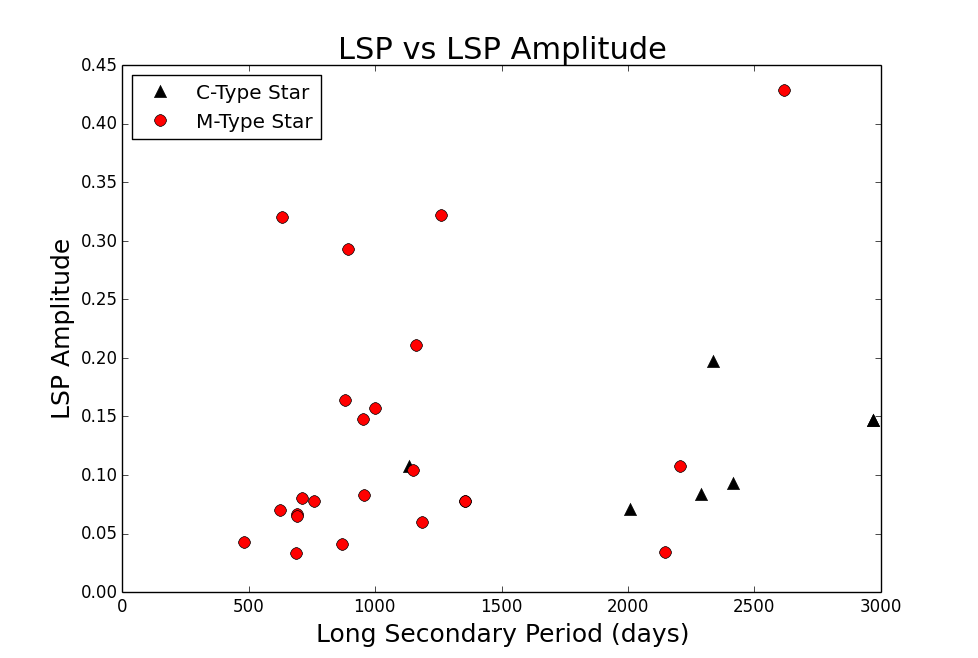}
\end{center}
\caption{The LSP amplitude, in magnitudes, as a function of LSP.  There is much scatter,
but the median amplitude
is about 0.10 mag, and there is no obvious correlation between the two.}
\end{figure}

\begin{figure}
\begin{center}
\includegraphics[height=7cm]{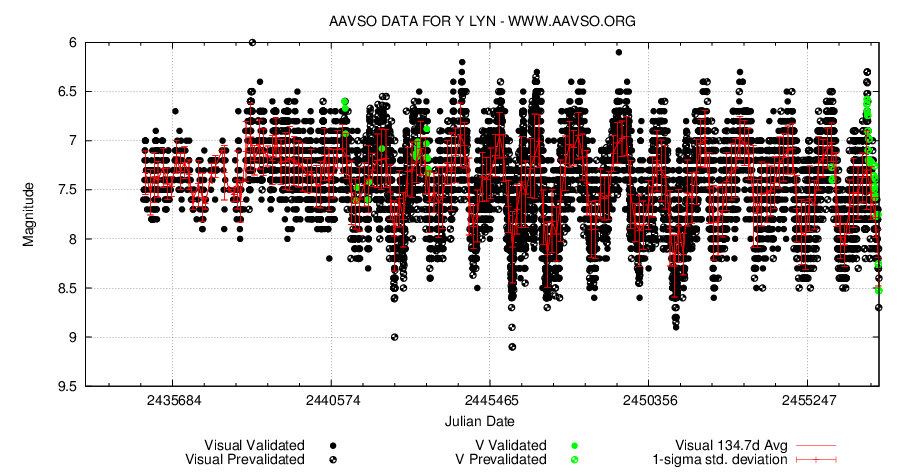}
\end{center}
\caption{The long-term light curve of Y Lyn, also showing the data averaged
in bins of 134.7 days (the pulsation period) to show the sawtoothed shape
of the LSP light curve (red curve) more clearly.  The LSP, as determined from VSTAR, is
1258.7 days, and this is consistent with the value which would be obtained
by counting cycles in this figure.  We have plotted all the observations, to
show how the density of the observations and the clarity of the LSP increase after JD 2440000.  The black points are visual; the green ones are photoelectric V.}
\end{figure}
\bigskip
 
\begin{figure}
\begin{center}
\includegraphics[height=7cm]{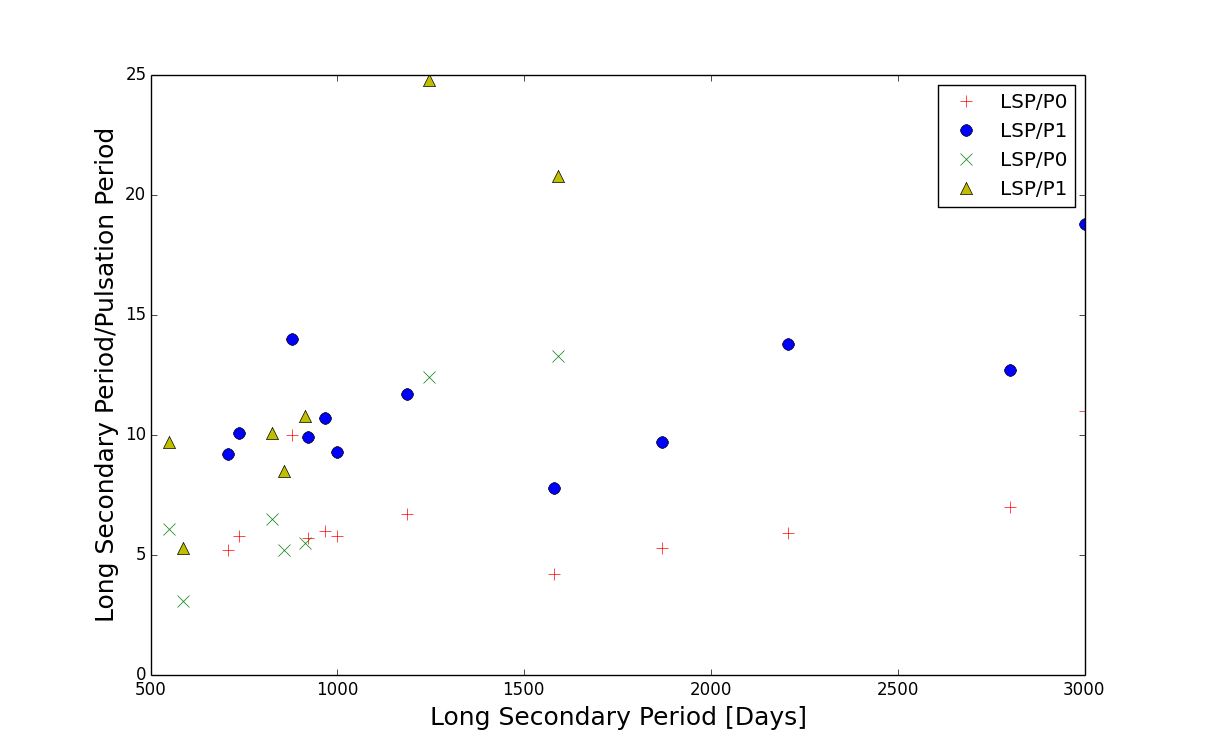}
\end{center}
\caption{For stars with both P0 and P1 excited: the ratio of the LSP
to P0 (plus signs and x's), and to P1 (filled circles and triangles).  The plus
signs and filled circles are from the visual data of Kiss {\it et al.} (1999)
and from Percy and Huang (2015); the x's and filled triangles are from
the ASAS data of Fuentes-Morales and Vogt (2014).} 
\end{figure}

\bigskip
\begin{figure}
\begin{center}
\includegraphics[height=7cm]{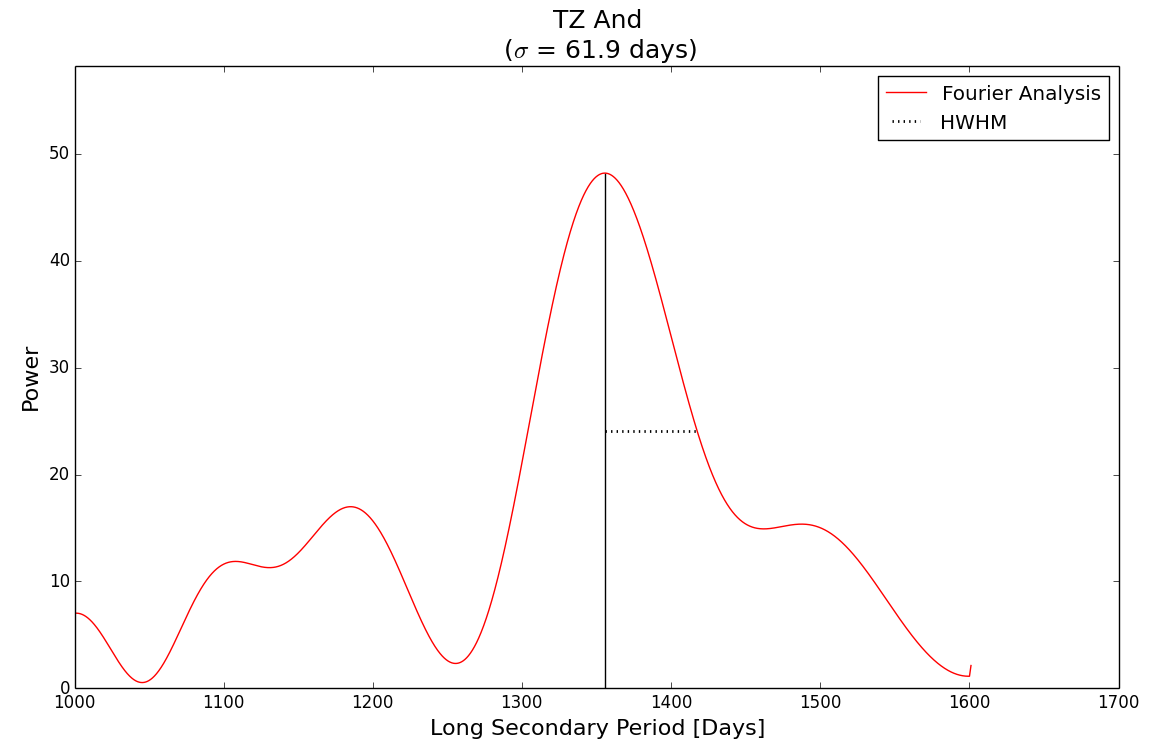}
\end{center}
\caption{The Fourier spectrum of TZ And, showing how the uncertainty $\sigma$ in
the LSP is determined from the half-width at half-maximum of the LSP peak.}
\end{figure}

\begin{figure}
\begin{center}
\includegraphics[height=7cm]{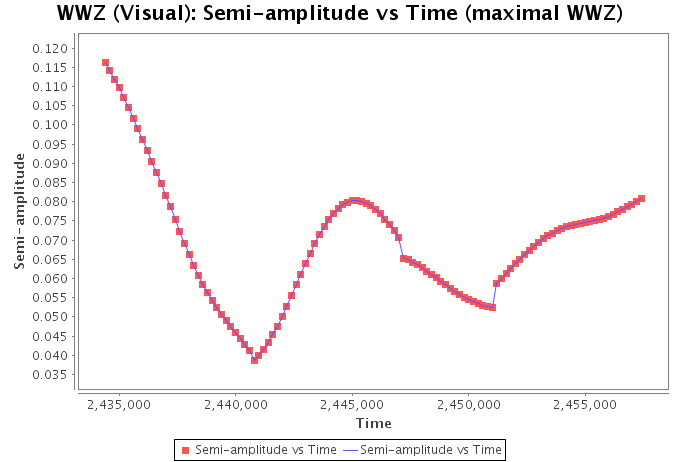}
\end{center}
\caption{The LSP amplitude of RT Cnc as a function of time, determined by wavelet analysis.  The amplitude
rises and falls on a time scale of about 20 LSPs.}
\end{figure}

\begin{figure}
\begin{center}
\includegraphics[height=7cm]{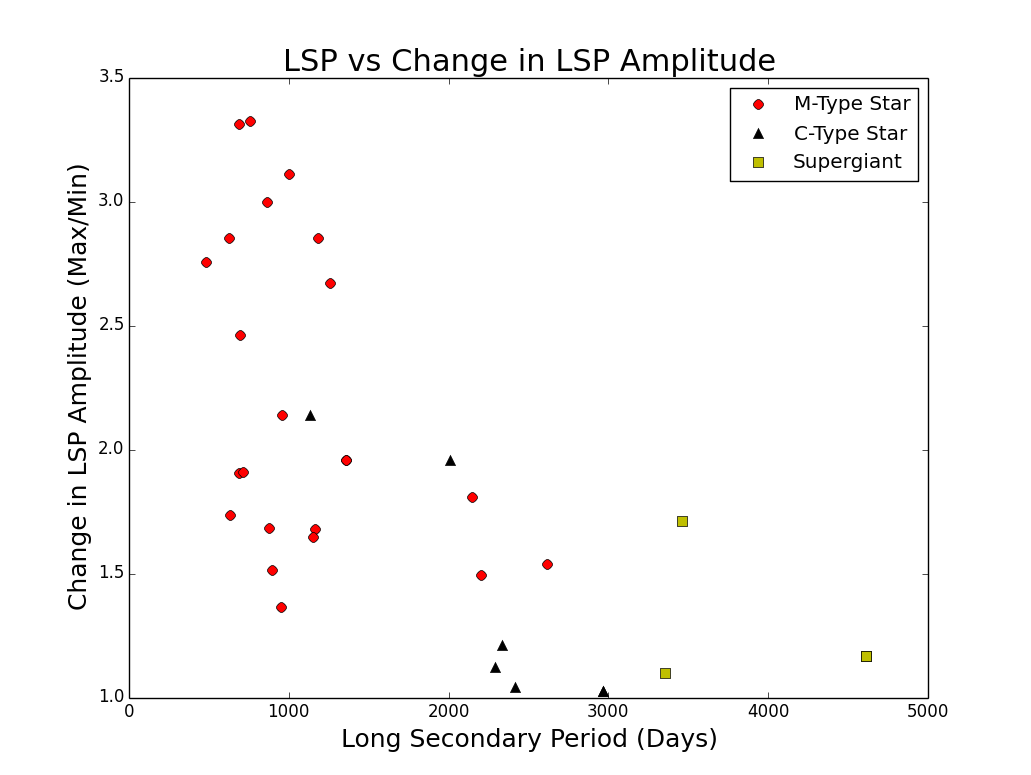}
\end{center}
\caption{The relative change in LSP amplitude (max/min) over the timespan
of the data, as a function of LSP.  As explained in the text, this
suggests that the LSP amplitude varies slowly, on a time scale of a
few tens of LSPs.}
\end{figure}

\begin{figure}
\begin{center}
\includegraphics[height=10cm]{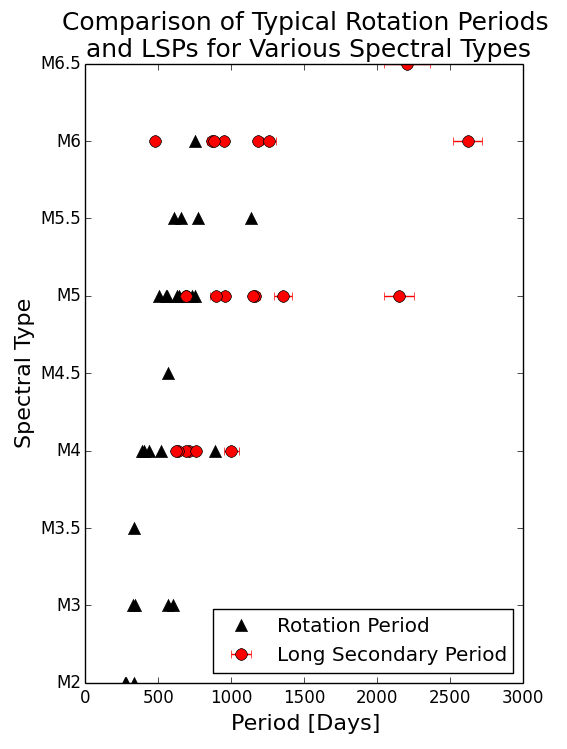}
\end{center}
\caption{The LSP and the rotation period, as a function of spectral type, for red giants.
See text for discussion.}
\end{figure}

\bigskip

\noindent
{\bf 4. Discussion}

\smallskip

Numerous possible explanations for the LSPs have been proposed and examined
but, as NWCS have described, most have not been successful.  These include radial or
non-radial pulsation -- unless there is some non-standard pulsation mode yet
to be discovered.  Standard binary models fail to explain the highly non-random distribution of 
velocity-curve amplitudes, shapes, and orientations (but the situation is reminiscent of the
``Barr effect" (Percy 2015) in which the velocity curves of spectroscopic
binary stars were distorted by mass transfer).

Saio {\it et al.} (2015) have proposed that oscillatory convective modes are
a possible explanation for the LSPs.  The timescales are consistent with
the LSPs, assuming that the modes are actually excited.  The amplitudes,
shapes, and color dependences of the phase curves are in reasonable agreement
with observations.  We would need some explanation for the variable
amplitudes of the LSPs.  Saio {\it et al.} (2015) state that there is a
minimum luminosity for the presence of oscillatory convective modes.  We
find LSPs in many less-luminous stars, with amplitudes comparable to those
in more luminous ones. 

Another interesting and still-viable explanation is the rise and fall of
giant convective cells, which are known to occur in cool stars (Stothers 2010).
The time scales of rise and fall are similar to those of LSPs, but we must
remember that these stars are also rotating, so this phenomena would be 
modulated by the rotation period.  We do not find evidence for two different
LSPs in the same star (section 3.6).  Also: we would need an explanation
for the variable amplitude of the LSPs.

Another possible explanation is the effect of a dusty cloud orbiting the red giant
with a low-mass companion in a close, circular orbit (Soszy\'{n}ski and Udalski
2014).  The period of variability would be the orbital period.  The authors
suggest that the red giant would be co-rotating with the companion, so the
periods would be comparable with those in Figure 9.  The
period would be a year or several, and would increase with the size of the star.
The phase curves (their Figures 5 and 6) are close to sinusoidal, except for
the one for i = 90 degrees, which is vaguely sawtooth, and also has the
largest amplitude.  We find sawtooth phase curves in two stars with large phase-curve amplitudes.
The amplitude of the LSP might change if the dusty cloud was periodically
depleted and replenished.
It seems rather improbable that a third of red giants should have such
configurations, though we should remember that the not-uncommon LSPs in RV Tauri 
stars (the so-called RVb stars) are thought to be due to the effect of a binary companion (Percy 1993, Waelkens and Waters 1993).  Derekas {\it et al.} (2006) noted that the period-luminosity relation for LSPs overlapped with that for known (eclipsing) binary stars,
which was an argument for a binary explanation.  Soszy\'{n}ski and
Udalski (2014) determined the shapes of the light curves which would result
from their model.  Our phase  curves do not seem to agree with theirs, though
ours are rather indistinct because of the lower accuracy of the visual
observations, and the presence of the pulsational variability.

Yet another explanation would be the presence of a large, hot/bright or cool/dark convective
cell, with a lifetime of a few tens of LSPs.  This would result in rotational
variability whose amplitude varied, as bright convective cells grew and
decayed.  (We note, however, that Stothers (2010) estimates the turnover
timescale as being closer to LSP than to 20-30 LSPs.)
The rotation periods of red giants can be estimated from their
projected rotational velocities (v sin i) (Zamanov {\it et al.} 2008)
and their radii (Zamanov {\it et al.} 2007) using the formula P = 50.6R/v
where P is the period in days, R is the radius in solar units, and v is the
equatorial rotational velocity in km/s.  For stars in the range M1III to
M6III, the rotational periods range mostly from 500 to 1000 days.  Figure 9 shows the
LSP and the rotation period of stars as a function of spectral type.
The rotation periods are estimated as described above.  They are highly
uncertain, as a result of the uncertainty in v sin i and R, and the unknown
value of sin i.  We note, however, that Olivier and Wood (2003) obtained
upper limits to rotation velocities in nearby stars with LSPs, and found
them to be too slow for rotation of a (single) spot to yield the observed
LSP values.  Figure 9 suggests that the rotation is too {\it fast}. 

The {\it general} similarity, however, between the rotation period and
the LSP, and the stability of the LSP is consistent with a rotational
origin for the LSP.  Whatever causes the LSP amplitude variations must
have a timescale which is proportional to LSP.

\medskip

\noindent
{\bf 5. Conclusions}

\smallskip

We have used systematic, sustained visual observations from the AAVSO
International Database, and the AAVSO time-series analysis software
package to study the LSPs in 27 red giants.  The LSPs are, on average,
8.1 $\pm$ 1.3 times the excited pulsation period, and the median visual amplitude is 0.1,
independent of the LSP.  The LSPs
are stable over
many decades, but the amplitudes rise and fall on a time scale of a few
tens of LSPs.  The LSP phase curves are not well-defined, but at least
two appear to have a sawtooth shape.  
The similarity between the LSP and the rotation period of the
star suggests that the LSP phenomenon may be due to rotational variability,
modulated by a large bright or dark convective cell, or by obscuring
material near the photosphere.  We also note that visual
observations, despite their lower accuracy, are able to delineate the
long-term behavior of these stars because they are sustained over up to a
century.   Whether or not the rotational hypothesis is correct, we believe
that we have added a few more ``knowns" to the NWCS list, and brought us a
bit closer to solving this longstanding mystery.
 
\medskip

\noindent
{\bf Acknowledgements}

\smallskip

We thank the AAVSO observers who made the observations on which this project
is based, the AAVSO staff who archived them and made them publicly available, and the developers
of the VSTAR package which we used for analysis.  We are very grateful to
Professor Peter Wood for reading and commenting on a draft of this
paper.   We acknowledge and thank the University
of Toronto Work-Study Program for financial support.  JRP thanks co-author ED,
an astronomy and physics (and English) major, for carrying out this project
so professionally.  ED thanks co-author JRP for the opportunity to work
on this project.
This project made
use of the SIMBAD database, maintained in Strasbourg, France.

\bigskip

\noindent
{\bf References}

\medskip

\noindent
Benn, D. 2013, VSTAR data analysis software (http://www.aavso.org/node/803).

\smallskip

\noindent
Derekas, A., Kiss, L.L., Bedding, T.R., Kjeldsen, H., Lah, P., and Szab\'{o}, Gy.M. 2006, {\it Astrophys. J. Letters}, {\bf 650}, L55.

\smallskip

\noindent
Eddington, A.S. and Plakidis, S. 1929, {\it Mon. Not. Roy. Astron. Soc.}, {\bf 90}, 65.

\smallskip

\noindent
Fuentes-Morales, I. and Vogt, N., 2014, {\it Astron. Nachr.}, {\bf 335}, 1072.

\smallskip

\noindent
Houk, N. 1963, {\it Astron. J.}, {\bf 68}, 253.

\smallskip

\noindent
Kafka, S. 2016, observations from the AAVSO International Database (https://www.aavso.org/aavso-international-database

\smallskip

\noindent
Kiss, L.L., Szatm\'{a}ry, K., Cadmus, R.R., Jr., and Mattei, J.A. 1999, {\it Astron. Astrophys.}, {\bf 346}, 542.

\smallskip

\noindent
Kiss, L.L., Szab\'{o}, G.Y. and Bedding, T.R. 2006, {\it Mon. Not. Roy. Astron. Soc.}, {\bf 372}, 1721.

\smallskip

\noindent
Knapp, G.R., Dobrovolsky, S.L., Ivesic, Z., Young, K., Crosas, M., Mattei, J.A., Rupen, M.P. 1999, {\it Astron. Astrophys.}, {\bf 351}, 97.

\smallskip

\noindent
Nicholls, C.P., Wood, P.R., Cioni, M.-R., and Soszy\'{n}ski, I. 2009, {\it Mon. Not. Roy. Astron. Soc.}, {\bf 399}, 2063 (NWCS).

\smallskip

\noindent
Nicholls, C.P., Wood, P.R. and Cioni, M.-R.L. 2010, {\it Mon. Not. Roy. Astron. Soc.}, {\bf 405}, 1770.

\smallskip

\noindent
O'Connell, D.J.K. 1933, {\it Bull. Harvard Obs.}, \#893, 19.

\smallskip

\noindent
Olivier, E.A. and Wood, P.R. 2003, {\it Astrophys. J.}, {\bf 584}, 1035.

\smallskip

\noindent
Payne-Gaposchkin, C. 1954, {\it Harvard Annals}. {\bf 113}, 189.

\smallskip

\noindent
Percy, J.R. 1993, in {\it Luminous High-Latitude Stars}, ed. D.D. Sasselov, ASP Conference Series 45, 295.

\smallskip

\noindent
Percy, J.R., Desjardins, A., Yu, L. and Landis, H.J. 1996, {\it Publ. Astron. Soc. Pacific}, {\bf 108}, 139.

\smallskip

\noindent
Percy, J.R. and Colivas, T. 1999, {\it Publ. Astron. Soc. Pacific}, {\bf 111}, 94.

\smallskip

\noindent
Percy, J.R., Wilson, J.B. and Henry, G.W. 2001, {\it Publ. Astron. Soc. Pacific}, {\bf 113}, 983.

\smallskip

\noindent
Percy, J.R. and Bakos, G.A. 2003, in {\it The Garrison Festschrift}, ed. Gray, R.O., Corbally, C. and Philip, A.G.D., L. Davis Press, 49.

\smallskip

\noindent
Percy, J.R., Mashintsova, M., Nasui, C.O., Seneviratne, R., and Henry, G.W. 2008, {\it Publ. Astron. Soc. Pacific}, {\bf 120}, 523.

\smallskip

\noindent
Percy, J.R. and Sato, H. 2009, {\it J. Roy. Astron. Soc. Canada}, {\bf 103}, 11.

\smallskip

\noindent
Percy, J.R. and Abachi, R. 2013, {\it J. Amer. Assoc. Var. Star Observers}, {\bf 41}, 1.

\smallskip

\noindent
Percy, J.R. and Yook, J.Y. 2014, {\it J. Amer. Assoc. Var. Star Observers}, {\bf 42}, 245.

\smallskip

\noindent
Percy, J.R. and Huang, D.J. 2015, {\it JAAVSO}, {\bf 43}, 118.

\smallskip

\noindent
Percy, J.R. 2015, {\it J. Roy. Astron. Soc. Canada}, {\bf 109}, 266.

\smallskip

\noindent
Price, S.D., Smith, B.J., Kuchar, T.A., Mizuno, D.R., and Kraemer, K.E. 2010,
{\it Astrophys. J. Suppl.}, {\bf 190}, 203.

\smallskip

\noindent
Saio, H., Wood, P.R., Takayama, M., and Ita, Y., 2015, {\it Mon. Not. Roy. Astron. Soc.}, {\bf 452}, 3863.

\smallskip

\noindent
Soszy\'{n}ski, I. and Udalski, A. 2014, {\it Astrophys. J.}, {\bf 788}, 13-6.

\smallskip

\noindent
Stothers, R.B. 2010, {\it Astrophys. J.}, {\bf 725}, 1170.

\smallskip

\noindent
Waelkens, C. and Waters, L.B.F.M. 1993, in {\it Luminous High-Latitude Stars}, ed. D.D. Sasselov, ASP Conference Series 45, 219.

\smallskip

\noindent
Wood, P.R. {\it et al.} 1999, in IAU Symp. \#191, {\it Asymptotic Giant
Branch Stars}, ed. T. Le Bertre, A. Lebre, and C. Waelkens, Cambridge:
Cambridge University Press, 151.

\smallskip

\noindent
Wood, P.R. 2000, {\it Publ. Astron. Soc. Australia}, {\bf 17}, 18.

\smallskip

\noindent
Yang, M. and Jiang, B.W. 2012, {\it Astrophys. J.}, {\bf 754}, 35.

\smallskip

\noindent
Zamanov, R.K., Bode, M.F., Melo, C.H.F., Bachev, R., Gomboc, A., Stateva, I.K., Porter, J.M., and Pritchard, J., 2007, {\it Mon. Not. Roy. Astron. Soc.}, {\bf 380},
1053.

\smallskip

\noindent
Zamanov, R.K., Bode, M.F., Melo, C.H.F., Stateva, I.K., Bachev, R., Gomboc, A., Konstantinova-Antova, R., and Stoyabov, K.A., 2008, {\it Mon. Not. Roy. Astron. Soc.}, {\bf 390},
377.

\smallskip

\begin{table}\small
\caption{Pulsation Periods and Long Secondary Periods in Red Giants}
\begin{tabular}{rrrrrrrr}
\hline
Star & SpT & JD(start) & P(d) & LSP & LSP/P & A/Amax/Amin & Notes \\
\hline
TZ And & M & 2440093 & 114.8 & 1355.1 & 11.8 & 0.078 0.096 0.049 & pc?, dsp \\
RZ Ari & M & all & 56.5 & 479.4 & 8.48 & 0.043 0.102 0.037 & pc? \\
T Ari & M & 2415000 & 320.6 & 2617.8 & 8.17 & 0.429 0.485 0.315 & pc?, * \\
RX Boo & M & all & 160.3 & 2205.1 & 13.76 & 0.108 0.136 0.091 & pc?, * \\
RS Cam & M & 2443500 & 90.5 & 999 & 11.04 & 0.157 0.255 0.082 & pc: \\
U Cam & C & 2425400 & 219.4 & 2967.4 & 13.53 & 0.147 0.151 0.147 & pcs \\
IX Car & SG & 2442500 & 371.5 & 4608.3 & 12.40 & 0.188 0.249 0.213 & pcs:, dsp \\
AA Cas & M & 2437500 & 80.1 & 866.8 & 10.82 & 0.041 0.066 0.022 & pc? \\
SS Cep & M & 2432500 & 101.1 & 955.2 & 9.45 & 0.083 0.137 0.064 & pcs: \\
RT Cnc & M & all & 89.3 & 691.7 & 7.75 & 0.067 0.116 0.035 & pc? \\
FS Com & M & 2440000 & 55.7 & 688.7 & 12.36 & 0.033 0.080 0.042 & pc? \\
Y CVn & C & 2430000 & 160 & 2008.9 & 6.88 & 0.071 0.096 0.049 & pcs, * \\
AW Cyg & C & 2431500 & 209: & 2289: & 6.40 & 0.084 0.091 0.081 & pcs:, * \\
BC Cyg & SG & 2437500 & 698.8 & 3459.6 & 4.95 & 0.128 0.171 0.010 & pc? \\
U Del & M & all & 119.0 & 1162.8 & 9.77 & 0.211 0.257 0.153 & pcs \\
RY Dra & C & 2432500 & 276.7: & 1135.6: & 4.10 & 0.108 0.137 0.064 & pcs, * \\
TX Dra & M & 2431702 & 77.5 & 711.8 & 9.18 & 0.080 0.193 0.101 & pcs:, dsp \\
RW Eri & M & all & 91.4 & 952.0 & 10.42 & 0.148 0.172 0.126 & pcs: \\
Z Eri & M & 2430000 & 78.5 & 692.4 & 8.82 & 0.065 0.197 0.080 & pc? \\
TU Gem & C & all & 214.6 & 2413.7 & 11.25 & 0.093 0.101 0.097 & pcs: \\
g Her & M & 2430000 & 87.6 & 878.7 & 10.03 & 0.164 0.222 0.132 & pcs, * \\
X Her & M & 2430240 & 176.6 & 1185.3 & 6.71 & 0.060 0.134 0.047 & pc? \\ 
V Hya & C & all & 531.4 & 6907.4 & 13.00 & 1.132 1.160 1.092 & * \\ 
RV Lac & M & all & 70.0 & 632.6 & 9.04 & 0.320 0.453 0.261 & pcst \\
Y Lyn & M & all & 134.7 & 1258.7 & 9.34 & 0.322 0.398 0.149 & pcst \\
W Ori & C & 2430000 & 210.7 & 2335.1 & 11.08 & 0.197 0.213 0.176 & pc? \\
SU Per & SG & 2432500 & 469.0 & 3355.7 & 7.16 & 0.115 0.122 0.111 & pc? \\
$\tau$4 Ser & M & all & 111 & 1151.2 & 13.51 & 0.104 0.117 0.071 & pcsa, *: \\
ST UMa & M & 2430000 & 90.3 & 623.1 & 6.90 & 0.070 0.117 0.041 & pc? \\
V UMi & M & 2430644 & 72.9 & 757.3 & 18.7 & 0.078 0.103 0.031 & pcs:, * \\
\hline
\end{tabular}
\end{table}

\end{document}